\documentclass[12pt]{article}
\usepackage{epsfig}
\usepackage{longtable}
\usepackage{amssymb}

\usepackage[colorlinks]{hyperref} 
\usepackage{xcolor}
\definecolor{lc}{rgb}{0.0,0.0,1.0} 
\hypersetup{colorlinks=true,urlcolor=lc,linkcolor=lc,citecolor=lc}

\newcommand{\br}{\bf \textcolor{red}}

\begin{document}

\setlength\LTcapwidth{\linewidth}

\begin{center}


{\Large \bf{Spontaneous double alpha decay: \\ First experimental limit and prospects of investigation}}

\vskip 1.0cm

{\bf V.I.~Tretyak}\footnote{tretyak@kinr.kiev.ua.}

\vskip 0.3cm

{\it Institute for Nuclear Research of NASU, 03028 Kyiv, Ukraine}

\end{center}

\vskip 0.5cm

\begin{abstract}

\noindent Nuclear decays with simultaneous emission of two alpha particles are energetically possible 
for a number of nuclides. Prospects of searching for such kind of decay for nuclides present in the 
natural isotopic composition of elements are discussed here. The first experimental limit on half-life 
for $2\alpha$ decay is set for $^{209}$Bi as $T_{1/2} > 2.9\times10^{20}$~y 
at 90\%~C.L., using the data of work [P. de Marcillac et al., Nature 422 (2003) 876]. 
Theoretical $T_{1/2}$ estimations for the process are also given. Using these values, 
which are on the level of $10^{33}$ y or more, one can conclude that the prospects of 
experimental observation of $2\alpha$ decay are very pessimistic. 

~


\end{abstract}

\section{Introduction}

We know today several kinds of decay of atomic nuclei in which two particles 
of the same nature are emitted simultaneously. 

Double beta ($2\beta$) decay, in which two electrons are emitted together with 
two antineutrinos, was predicted in 1935 by Goeppert-Mayer \cite{Goe35}
and is registered to-date for 14 nuclides \cite{Saa13,Bla20}.
Neutrinoless mode of this decay, in which only two electrons are emitted without 
antineutrinos, was first discussed by Furry in 1939 \cite{Fur39}.
It is forbidden in the Standard Model (SM) of particles because it violates 
the lepton number by two units, but is predicted by many theories which consider 
the SM as a low-energy approximation; it is not observed yet but is actively 
searched for in many worldwide experiments \cite{Bla20,Dol19}. 

Deexcitation of an excited nuclear state with emission of two gammas, $2\gamma$ decay, 
was considered by Goeppert-Mayer in 1929 \cite{Goe29} (see also \cite{Goe31}). 
It was observed in 1963 for nuclei with $0^+ \to 0^+$ transitions, like 
$^{16}$O, $^{40}$Ca and $^{90}$Zr, where emission of single $\gamma$ is forbidden, 
and allowed are emissions of a conversion electron, or electron-positron pair, 
or two gammas \cite{Sut63}. 
In 2015, it was also observed at the first time for $^{137}$Ba, where emission of 
a single $\gamma$ is allowed, as a competitive mode on the level of $10^{-6}$ 
\cite{Wal15}. 

Also, nuclear decays with simultaneous emission of two protons ($2p$) and two neutrons ($2n$) 
are known; see \cite{Pfu12} and Refs. therein.

However, possibility of nuclear decay with emission of two alpha particles at the same time 
($2\alpha$ decay) was not analyzed on the same level of attention as the $2\beta$, $2\gamma$, 
$2p$ and $2n$ radioactivities. 
While the idea on $2\alpha$ decay appeared independently, further scan of the literature showed 
that it was already discussed: to our knowledge, 
at the first time by Novikov in \cite{Nov79}, with estimation of half-lives for few nuclides in 
\cite{Ber88}. It was considered also in 
theoretical article \cite{Poe85} (see also \cite{Poe89}). There were no 
experimental searches for such processes yet. 

Here we analyze possible experimental investigations of nuclear $2\alpha$ decay (Section 2), 
give theoretical estimations of half-lives for a number of nuclides present in the 
natural isotopic composition of elements (Section 3) and set the first experimental 
limit for such a process (Section 4). 
It should be noted that in interactions of nuclei with high-energy particles 
the defragmentation of nuclei with emission of few $\alpha$ particles is possible 
\cite{Oer10} but we are interested here in the spontaneous $2\alpha$ decay.

\section{Possible experimental approaches to search for $2\alpha$ \\ decay}

Analyzing atomic masses from the last atomic mass evaluation \cite{Wan17}, 
one can see that $2\alpha$ decay is energetically possible for many nuclei 
(for 1459 isotopes from 3436 listed in \cite{Wan17}). Energy releases $Q_{2\alpha}$ 
can reach quite big values (maximal value is 22.7~MeV for $^{276}$Cn) that, 
supposing an exponential dependence of half-life $T_{1/2}$ on $Q_{2\alpha}$ 
similar to single alpha decay, increases probability of this process. 
However, many of such nuclides are unstable with short half-lives respective 
to their main decay branch; they should be created at accelerators or 
extracted from fission products at nuclear reactors. 
It would be quite difficult to see a rare $2\alpha$ process on the big 
radioactive $\alpha$ and $\beta$ backgrounds. 
Instead, we will concentrate here on techniques which are typical in searches for 
rare nuclear decays, like $2\beta$ \cite{Saa13,Bla20,Dol19}, 
single $\alpha$ and $\beta$ \cite{Bel19}, 
dark matter \cite{Ber18,Apr17}, or 
in neutrino measurements \cite{Ali09}.

In such an approach, massive (from $\sim 0.1$~kg to $\sim 1$~t) samples consisting of 
nuclides present in the natural isotopic composition of elements \cite{Mei16} are measured, 
usually in deep underground laboratories to decrease the background from cosmic muons. 
Detectors, massive shieldings and all surrounding details are built from super-pure 
materials to suppress natural radioactivity that is necessary to see a small useful signal. 
Nuclides present in the natural composition of elements which are potentially 
unstable in relation to $2\alpha$ decay are listed in Table~1.

\begin{footnotesize}
\begin{longtable}{llllll}
\caption{Naturally abundant 80 nuclides candidates for 2$\alpha$ decay. 
$\delta$ is the natural abundance of the mother isotope \cite{Mei16}. 
Energy release $Q_{2\alpha}$ is calculated with atomic masses \cite{Wan17}. 
Half-life $T_{1/2}$ is given for the daughter isotope \cite{Fir98}. 
If value of $Q_{2\alpha}$ is greater than 3~MeV, it is marked in bold (red); 
in this case big abundance (if so) and unstable daughter isotope (if so) 
are also marked in bold (red). 
In the last column, the theoretical estimation of $T_{1/2}$ is given, 
calculated with receipt from \cite{Poe02} for $^8$Be emission. 
$2\varepsilon$ is for double electron capture,
$\varepsilon\beta^+$ is for electron capture with positron emission.} \\
\hline
Mother         & $\delta$,\%       & $Q_{2\alpha}$, keV    & Daughter                 & Additional possible $\alpha$ and $\beta$ decay                              & Theor.              \\
isotope        & \cite{Mei16}      & \cite{Wan17}          & isotope                  & modes of mother isotope                                                     & $T_{1/2}$, y        \\
~              & ~                 & ~                     & ($T_{1/2}$ \cite{Fir98}) & ~                                                                           & ~                   \\
\hline         
\endfirsthead
\multicolumn{6}{r}{\it{Table 1 continued}} \\
\hline
Mother         & $\delta$,\%       & $Q_{2\alpha}$, keV    & Daughter                 & Additional possible $\alpha$ and $\beta$ decay                              & Theor.              \\
isotope        & \cite{Mei16}      & \cite{Wan17}          & isotope                  & modes of mother isotope                                                     & $T_{1/2}$, y        \\
~              & ~                 & ~                     & ($T_{1/2}$ \cite{Fir98}) & ~                                                                           & ~                   \\
\hline         
~              & ~                 & ~                     & ~                        & ~                                                                           & ~                   \\ 
\endhead
~              & ~                 & ~                     & ~                        & ~                                                                           & ~                   \\ 
\hline \multicolumn{6}{r}{\it{Continued on next page}}
\endfoot
\endlastfoot
~              & ~                 & ~                     & ~                        & ~                                                                           & ~                   \\ 
$^{144}$Nd     & 23.798            & 289.1  $\pm$ 1.4      & $^{136}$Ba (stable)      & $\alpha$ ($^{144}$Nd $\to$ $^{140}$Ce $\nrightarrow$ $^{136}$Ba)            & ~                   \\  
$^{145}$Nd     & 8.293             & 1439.5 $\pm$ 1.4      & $^{137}$Ba (stable)      & $\alpha$ ($^{145}$Nd $\to$ $^{141}$Ce $\nrightarrow$ $^{137}$Ba)            & ~                   \\  
$^{146}$Nd     & 17.189            & 2485.9 $\pm$ 1.4      & $^{138}$Ba (stable)      & 2$\beta^-$, $\alpha$ ($^{146}$Nd $\to$ $^{142}$Ce $\to$ $^{138}$Ba)         & ~                   \\ 
$^{148}$Nd     & 5.756             & 1011.5 $\pm$ 8.2      & $^{140}$Ba (12.7 d)      & 2$\beta^-$, $\alpha$ ($^{148}$Nd $\to$ $^{144}$Ce $\to$ $^{140}$Ba)         & ~                   \\  
~              & ~                 & ~                     & ~                        & ~                                                                           & ~                   \\ 
$^{147}$Sm     & 15.00             & 2831.7 $\pm$ 7.5      & $^{139}$Ce (137.6 d)     & $\alpha$ ($^{147}$Sm $\to$ $^{143}$Nd $\to$ $^{139}$Ce)                     & ~                   \\  
$^{148}$Sm     & 11.25             & \br{3890.0 $\pm$ 2.1} & $^{140}$Ce (stable)      & $\alpha$ ($^{148}$Sm $\to$ $^{144}$Nd $\to$ $^{140}$Ce)                     & $1.3\times10^{58}$  \\  
$^{149}$Sm     & \br{13.82}        & \br{3447.3 $\pm$ 2.1} & \br{$^{141}$Ce (32.5 d)} & $\alpha$ ($^{149}$Sm $\to$ $^{145}$Nd $\to$ $^{141}$Ce)                     & $4.7\times10^{66}$  \\  
$^{150}$Sm     & 7.37              & 2632.2 $\pm$ 2.8      & $^{142}$Ce (stable)      & $\alpha$ ($^{150}$Sm $\to$ $^{146}$Nd $\to$ $^{142}$Ce)                     & ~                   \\  
$^{152}$Sm     & 26.74             & 819.4 $\pm$ 3.1       & $^{144}$Ce (284.9 d)     & $\alpha$ ($^{152}$Sm $\to$ $^{148}$Nd $\to$ $^{144}$Ce)                     & ~                   \\  
~              & ~                 & ~                     & ~                        & ~                                                                           & ~                   \\ 
$^{151}$Eu     & \br{47.81}        & \br{3565.6 $\pm$ 2.3} & \br{$^{143}$Pr (13.6 d)} & $\alpha$ ($^{151}$Eu $\to$ $^{147}$Pm $\to$ $^{143}$Pr)                     & $1.5\times10^{66}$  \\  
$^{153}$Eu     & 52.191            & 1408.9 $\pm$ 7.3      & $^{145}$Pr (6.0 h)       & $\alpha$ ($^{153}$Eu $\to$ $^{149}$Pm $\to$ $^{145}$Pr)                     & ~                   \\  
~              & ~                 & ~                     & ~                        & ~                                                                           & ~                   \\ 
$^{152}$Gd     & 0.20              & \br{4191.2 $\pm$ 1.8} & $^{144}$Nd (2.3e15 y)    & 2$\varepsilon, \alpha$ ($^{152}$Gd $\to$ $^{148}$Sm $\to$ $^{144}$Nd)       & $4.1\times10^{56}$  \\  
$^{154}$Gd     & 2.18              & 2370.1 $\pm$ 1.8      & $^{146}$Nd (stable)      & $\alpha$ ($^{154}$Gd $\to$ $^{150}$Sm $\to$ $^{146}$Nd)                     & ~                   \\  
$^{155}$Gd     & 14.80             & 1227.0 $\pm$ 1.8      & $^{147}$Nd (11.0 d)      & $\alpha$ ($^{155}$Gd $\to$ $^{151}$Sm $\to$ $^{147}$Nd)                     & ~                   \\  
$^{156}$Gd     & 20.47             & 23.3 $\pm$ 2.4        & $^{148}$Nd (stable)      & $\alpha$ ($^{156}$Gd $\to$ $^{152}$Sm $\to$ $^{148}$Nd)                     & ~                   \\  
~              & ~                 & ~                     & ~                        & ~                                                                           & ~                   \\ 
$^{156}$Dy     & 0.056             & \br{3957.5 $\pm$ 1.8} & $^{148}$Sm (7.0e15 y)    & $\varepsilon\beta^+, \alpha$ ($^{156}$Dy $\to$ $^{152}$Gd $\to$ $^{148}$Sm) & $1.9\times10^{64}$  \\  
$^{158}$Dy     & 0.095             & 1794.0 $\pm$ 2.7      & $^{150}$Sm (stable)      & 2$\varepsilon$, $\alpha$ ($^{158}$Dy $\to$ $^{154}$Gd $\to$ $^{150}$Sm)     & ~                   \\  
$^{160}$Dy     & 2.329             & 240.0 $\pm$ 1.4       & $^{152}$Sm (stable)      & $\alpha$ ($^{160}$Dy $\to$ $^{156}$Gd $\nrightarrow$ $^{152}$Sm)            & ~                   \\  
~              & ~                 & ~                     & ~                        & ~                                                                           & ~                   \\ 
$^{162}$Er     & 0.139             & 2521.6 $\pm$ 1.5      & $^{154}$Gd (stable)      & $\varepsilon\beta^+, \alpha$ ($^{162}$Er $\to$ $^{158}$Dy $\to$ $^{154}$Gd) & ~                   \\  
$^{164}$Er     & 1.601             & 1742.2 $\pm$ 1.4      & $^{156}$Gd (stable)      & 2$\epsilon$, $\alpha$ ($^{164}$Er $\to$ $^{160}$Dy $\to$ $^{156}$Gd)        & ~                   \\  
$^{166}$Er     & 33.503            & 913.7 $\pm$ 1.7       & $^{158}$Gd (stable)      & $\alpha$ ($^{166}$Er $\to$ $^{162}$Dy $\to$ $^{158}$Gd)                     & ~                   \\  
$^{167}$Er     & 22.869            & 420.4 $\pm$ 1.7       & $^{159}$Gd (18.5 h)      & $\alpha$ ($^{167}$Er $\to$ $^{163}$Dy $\nrightarrow$ $^{159}$Gd)            & ~                   \\  
$^{168}$Er     & 26.978            & 100.7 $\pm$ 1.7       & $^{160}$Gd (stable)      & $\alpha$ ($^{168}$Er $\to$ $^{164}$Dy $\nrightarrow$ $^{160}$Gd)            & ~                   \\  
~              & ~                 & ~                     & ~                        & ~                                                                           & ~                   \\ 
$^{169}$Tm     & 100               & 1336.5 $\pm$ 1.6      & $^{161}$Tb (6.9 d)       & $\alpha$ ($^{169}$Tm $\to$ $^{165}$Ho $\to$ $^{161}$Tb)                     & ~                   \\  
~              & ~                 & ~                     & ~                        & ~                                                                           & ~                   \\ 
$^{168}$Yb     & 0.123             & \br{3241.0 $\pm$ 1.4} & $^{160}$Dy (stable)      & $\varepsilon\beta^+, \alpha$ ($^{168}$Yb $\to$ $^{164}$Er $\to$ $^{160}$Dy) & ~                   \\  
$^{170}$Yb     & 2.982             & 2567.7 $\pm$ 0.8      & $^{162}$Dy (stable)      & $\alpha$ ($^{170}$Yb $\to$ $^{166}$Er $\to$ $^{162}$Dy)                     & ~                   \\  
$^{171}$Yb     & 14.086            & 2224.5 $\pm$ 0.8      & $^{163}$Dy (stable)      & $\alpha$ ($^{171}$Yb $\to$ $^{167}$Er $\to$ $^{163}$Dy)                     & ~                   \\  
$^{172}$Yb     & 21.686            & 1862.7 $\pm$ 0.8      & $^{164}$Dy (stable)      & $\alpha$ ($^{172}$Yb $\to$ $^{168}$Er $\to$ $^{164}$Dy)                     & ~                   \\  
$^{173}$Yb     & 16.103            & 1211.6 $\pm$ 0.8      & $^{165}$Dy (2.3 h)       & $\alpha$ ($^{173}$Yb $\to$ $^{169}$Er $\to$ $^{165}$Dy)                     & ~                   \\  
$^{174}$Yb     & 32.025            & 790.4 $\pm$ 0.9       & $^{166}$Dy (81.6 h)      & $\alpha$ ($^{174}$Yb $\to$ $^{170}$Er $\to$ $^{166}$Dy)                     & ~                   \\  
$^{176}$Yb     & 12.995            & 218 $\pm$ 140         & $^{168}$Dy (8.7 m)       & 2$\beta$, $\alpha$ ($^{176}$Yb $\to$ $^{172}$Er $\nrightarrow$ $^{168}$Dy)  & ~                   \\  
~              & ~                 & ~                     & ~                        & ~                                                                           & ~                   \\ 
$^{175}$Lu     & 97.401            & 2265.2 $\pm$ 5.4      & $^{167}$Ho (3.1 h)       & $\alpha$ ($^{175}$Lu $\to$ $^{171}$Tm $\to$ $^{167}$Ho)                     & ~                   \\  
$^{176}$Lu     & 2.599             & 1829 $\pm$ 30         & $^{168}$Ho (3.0 m)       & $\alpha$ ($^{176}$Lu $\to$ $^{172}$Tm $\to$ $^{168}$Ho)                     & ~                   \\  
~              & ~                 & ~                     & ~                        & ~                                                                           & ~                   \\ 
$^{174}$Hf     & 0.16              & \br{4231.7 $\pm$ 2.5} & $^{166}$Er (stable)      & $\varepsilon\beta^+, \alpha$ ($^{174}$Hf $\to$ $^{170}$Yb $\to$ $^{166}$Er) & $8.2\times10^{69}$  \\  
$^{176}$Hf     & 5.26              & \br{3565.0 $\pm$ 1.9} & $^{168}$Er (stable)      & $\alpha$ ($^{176}$Hf $\to$ $^{172}$Yb $\to$ $^{168}$Er)                     & ~                   \\  
$^{177}$Hf     & \br{18.60}        & \br{3192.7 $\pm$ 1.8} & \br{$^{169}$Er (9.4 d)}  & $\alpha$ ($^{177}$Hf $\to$ $^{173}$Yb $\to$ $^{169}$Er)                     & $2.4\times10^{93}$  \\  
$^{178}$Hf     & 27.28             & 2823.6 $\pm$ 2.1      & $^{170}$Er (stable)      & $\alpha$ ($^{178}$Hf $\to$ $^{174}$Yb $\to$ $^{170}$Er)                     & ~                   \\  
$^{179}$Hf     & 13.62             & 2406.2 $\pm$ 2.1      & $^{171}$Er (7.5 h)       & $\alpha$ ($^{179}$Hf $\to$ $^{175}$Yb $\to$ $^{171}$Er)                     & ~                   \\  
$^{180}$Hf     & 35.08             & 1854.4 $\pm$ 4.3      & $^{172}$Er (49.3 h)      & $\alpha$ ($^{180}$Hf $\to$ $^{176}$Yb $\to$ $^{172}$Er)                     & ~                   \\  
~              & ~                 & ~                     & ~                        & ~                                                                           & ~                   \\ 
$^{180}$Ta     & 0.01201           & \br{3591.7 $\pm$ 5.8} & \br{$^{172}$Tm (63.6 h)} & $\alpha$ ($^{180}$Ta $\to$ $^{176}$Lu $\to$ $^{172}$Tm)                     & $1.2\times10^{85}$  \\  
$^{181}$Ta     & 99.98799          & 2967.9 $\pm$ 4.6      & $^{173}$Tm (8.2 h)       & $\alpha$ ($^{181}$Ta $\to$ $^{177}$Lu $\to$ $^{173}$Tm)                     & ~                   \\  
~              & ~                 & ~                     & ~                        & ~                                                                           & ~                   \\ 
$^{180}$W      & 0.12              & \br{4769.5 $\pm$ 1.4} & $^{172}$Yb (stable)      & 2$\varepsilon$, $\alpha$ ($^{180}$W $\to$ $^{176}$Hf $\to$ $^{172}$Yb)      & $1.4\times10^{64}$  \\  
$^{182}$W      & 26.50             & \br{3848.6 $\pm$ 0.7} & $^{174}$Yb (stable)      & $\alpha$ ($^{182}$W $\to$ $^{178}$Hf $\to$ $^{174}$Yb)                      & ~                   \\  
$^{183}$W      & \br{14.31}        & \br{3480.1 $\pm$ 0.7} & \br{$^{175}$Yb (4.2 d)}  & $\alpha$ ($^{183}$W $\to$ $^{179}$Hf $\to$ $^{175}$Yb)                      & $6.4\times10^{89}$  \\  
$^{184}$W      & 30.64             & 2936.1 $\pm$ 0.7      & $^{176}$Yb (stable)      & $\alpha$ ($^{184}$W $\to$ $^{180}$Hf $\to$ $^{176}$Yb)                      & ~                   \\  
$^{186}$W      & 28.43             & 2337 $\pm$ 10         & $^{178}$Yb (74 m)        & 2$\beta^-$, $\alpha$ ($^{186}$W $\to$ $^{182}$Hf $\to$ $^{178}$Yb)          & ~                   \\  
~              & ~                 & ~                     & ~                        & ~                                                                           & ~                   \\ 
$^{185}$Re     & \br{37.40}        & \br{3714.9 $\pm$ 1.5} & \br{$^{177}$Lu (6.7 d)}  & $\alpha$ ($^{185}$Re $\to$ $^{181}$Ta $\to$ $^{177}$Lu)                     & $1.2\times10^{86}$  \\  
$^{187}$Re     & \br{62.60}        & \br{2992.6 $\pm$ 5.2} & \br{$^{179}$Lu (4.6 h)}  & $\alpha$ ($^{187}$Re $\to$ $^{183}$Ta $\to$ $^{179}$Lu)                     & $7.9\times10^{105}$ \\  
~              & ~                 & ~                     & ~                        & ~                                                                           & ~                   \\ 
$^{184}$Os     & 0.02              & \br{5474.0 $\pm$ 1.7} & $^{176}$Hf (stable)      & $\varepsilon\beta^+, \alpha$ ($^{184}$Os $\to$ $^{180}$W $\to$ $^{176}$Hf)  & $1.4\times10^{57}$  \\  
$^{186}$Os     & 1.59              & \br{4285.5 $\pm$ 1.6} & $^{178}$Hf (stable)      & $\alpha$ ($^{186}$Os $\to$ $^{182}$W $\to$ $^{178}$Hf)                      & $8.0\times10^{75}$  \\  
$^{187}$Os     & 1.96              & \br{4394.1 $\pm$ 1.6} & $^{179}$Hf (stable)      & $\alpha$ ($^{187}$Os $\to$ $^{183}$W $\to$ $^{179}$Hf)                      & $6.9\times10^{73}$  \\  
$^{188}$Os     & 13.24             & \br{3792.3 $\pm$ 1.6} & $^{180}$Hf (stable)      & $\alpha$ ($^{188}$Os $\to$ $^{184}$W $\to$ $^{180}$Hf)                      & ~                   \\  
$^{189}$Os     & \br{16.15}        & \br{3566.3 $\pm$ 1.6} & \br{$^{181}$Hf (42.4 d)} & $\alpha$ ($^{189}$Os $\to$ $^{185}$W $\to$ $^{181}$Hf)                      & $4.0\times10^{91}$  \\  
$^{190}$Os     & 26.26             & 2491.9 $\pm$ 6.2      & $^{182}$Hf (9.0e5 y)     & $\alpha$ ($^{190}$Os $\to$ $^{186}$W $\to$ $^{182}$Hf)                      & ~                   \\  
$^{192}$Os     & 40.78             & 767 $\pm$ 40          & $^{184}$Hf (4.1 h)       & 2$\beta^-$, $\alpha$ ($^{192}$Os $\to$ $^{188}$W $\to$ $^{184}$Hf)          & ~                   \\  
~              & ~                 & ~                     & ~                        & ~                                                                           & ~                   \\ 
$^{191}$Ir     & \br{37.3}         & \br{3734.2 $\pm$ 1.9} & \br{$^{183}$Ta (5.1 d)}  & $\alpha$ ($^{191}$Ir $\to$ $^{187}$Re $\to$ $^{183}$Ta)                     & $2.8\times10^{89}$  \\  
$^{193}$Ir     & 62.7              & 2008 $\pm$ 14         & $^{185}$Ta (49.4 m)      & $\alpha$ ($^{193}$Ir $\to$ $^{189}$Re $\to$ $^{185}$Ta)                     & ~                   \\  
~              & ~                 & ~                     & ~                        & ~                                                                           & ~                   \\ 
$^{190}$Pt     & 0.012             & \br{6089.8 $\pm$ 1.0} & $^{182}$W (stable)       & $\varepsilon\beta^+, \alpha$ ($^{190}$Pt $\to$ $^{186}$Os $\to$ $^{182}$W)  & $2.7\times10^{52}$  \\  
$^{192}$Pt     & 0.782             & \br{4567.1 $\pm$ 2.7} & $^{184}$W (stable)       & $\alpha$ ($^{192}$Pt $\to$ $^{188}$Os $\to$ $^{184}$W)                      & $1.4\times10^{74}$  \\  
$^{194}$Pt     & 32.864            & 2898.6 $\pm$ 1.3      & $^{186}$W (stable)       & $\alpha$ ($^{194}$Pt $\to$ $^{190}$Os $\to$ $^{186}$W)                      & ~                   \\  
$^{195}$Pt     & 33.775            & 2263.3 $\pm$ 1.3      & $^{187}$W (23.7 h)       & $\alpha$ ($^{195}$Pt $\to$ $^{191}$Os $\to$ $^{187}$W)                      & ~                   \\  
$^{196}$Pt     & 25.211            & 1173.5 $\pm$ 3.1      & $^{188}$W (69.4 d)       & $\alpha$ ($^{196}$Pt $\to$ $^{192}$Os $\to$ $^{188}$W)                      & ~                   \\  
~              & ~                 & ~                     & ~                        & ~                                                                           & ~                   \\ 
$^{197}$Au     & 100               & 1989.5 $\pm$ 8.2      & $^{189}$Re (24.3 h)      & $\alpha$ ($^{197}$Au $\to$ $^{193}$Ir $\to$ $^{189}$Re)                     & ~                   \\  
~              & ~                 & ~                     & ~                        & ~                                                                           & ~                   \\ 
$^{196}$Hg     & 0.15              & \br{4461.5 $\pm$ 3.0} & $^{188}$Os (stable)      & 2$\varepsilon$, $\alpha$ ($^{196}$Hg $\to$ $^{192}$Pt $\to$ $^{188}$Os)     & $3.4\times10^{79}$  \\  
$^{198}$Hg     & 10.04             & 2903.6 $\pm$ 0.8      & $^{190}$Os (stable)      & $\alpha$ ($^{198}$Hg $\to$ $^{194}$Pt $\to$ $^{190}$Os)                     & ~                   \\  
$^{199}$Hg     & 16.94             & 1999.3 $\pm$ 0.8      & $^{191}$Os (15.4 d)      & $\alpha$ ($^{199}$Hg $\to$ $^{195}$Pt $\to$ $^{191}$Os)                     & ~                   \\  
$^{200}$Hg     & 23.14             & 1529.1 $\pm$ 2.4      & $^{192}$Os (stable)      & $\alpha$ ($^{200}$Hg $\to$ $^{196}$Pt $\to$ $^{192}$Os)                     & ~                   \\  
$^{201}$Hg     & 13.17             & 881.9 $\pm$ 2.4       & $^{193}$Os (30.5 h)      & $\alpha$ ($^{201}$Hg $\to$ $^{197}$Pt $\to$ $^{193}$Os)                     & ~                   \\  
$^{202}$Hg     & 29.74             & 240.0 $\pm$ 2.5       & $^{194}$Os (6.0 y)       & $\alpha$ ($^{202}$Hg $\to$ $^{198}$Pt $\to$ $^{194}$Os)                     & ~                   \\  
~              & ~                 & ~                     & ~                        & ~                                                                           & ~                   \\ 
$^{203}$Tl     & 29.44             & 1081.0 $\pm$ 1.8      & $^{195}$Ir (2.5 h)       & $\alpha$ ($^{203}$Tl $\to$ $^{199}$Au $\to$ $^{195}$Ir)                     & ~                   \\  
~              & ~                 & ~                     & ~                        & ~                                                                           & ~                   \\ 
$^{204}$Pb     & 1.4               & 2684.8 $\pm$ 1.3      & $^{196}$Pt (stable)      & $\alpha$ ($^{204}$Pb $\to$ $^{200}$Hg $\to$ $^{196}$Pt)                     & ~                   \\  
$^{206}$Pb     & 24.1              & 1268.6 $\pm$ 2.4      & $^{198}$Pt (stable)      & $\alpha$ ($^{206}$Pb $\to$ $^{202}$Hg $\to$ $^{196}$Pt)                     & ~                   \\  
$^{207}$Pb     & 22.1              & 86.8 $\pm$ 2.4        & $^{199}$Pt (30.8 m)      & $\alpha$ ($^{207}$Pb $\to$ $^{203}$Hg $\nrightarrow$ $^{199}$Pt)            & ~                   \\  
$^{208}$Pb     & 52.4              & 0.7 $\pm$ 20          & $^{200}$Pt (12.5 h)      & $\alpha$ ($^{208}$Pb $\to$ $^{204}$Hg $\nrightarrow$ $^{200}$Pt)            & ~                   \\  
~              & ~                 & ~                     & ~                        & ~                                                                           & ~                   \\ 
$^{209}$Bi     & \br{100}          & \br{3292.2 $\pm$ 3.5} & \br{$^{201}$Au (26 m)}   & $\alpha$ ($^{209}$Bi $\to$ $^{205}$Tl $\to$ $^{201}$Au)                     & $4.3\times10^{113}$ \\  
~              & ~                 & ~                     & ~                        & ~                                                                           & ~                   \\ 
$^{232}$Th     & \br{99.98}        & \br{8151.9 $\pm$ 9.9} & \br{$^{224}$Rn (107 m)}  & 2$\beta^-$, $\alpha$ ($^{232}$Th $\to$ $^{228}$Ra $\to$ $^{224}$Rn)         & $6.3\times10^{46}$  \\  
~              & ~                 & ~                     & ~                        & ~                                                                           & ~                   \\ 
$^{231}$Pa$^*$ & \br{100}          & \br{10192.2 $\pm$ 2.6}& \br{$^{223}$Fr (21.8 m)} & $\alpha$ ($^{231}$Pa $\to$ $^{227}$Ac $\to$ $^{223}$Fr)                     & $1.5\times10^{33}$  \\  
~              & ~                 & ~                     & ~                        & ~                                                                           & ~                   \\ 
$^{234}$U      & 0.0054            & \br{9627.4 $\pm$ 2.2} & $^{226}$Ra (1600 y)      & $\alpha$ ($^{234}$U $\to$ $^{230}$Th $\to$ $^{226}$Ra)                      & $6.3\times10^{37}$  \\  
$^{235}$U      & \br{0.7204}       & \br{8891.3 $\pm$ 2.2} & \br{$^{227}$Ra (42.2 m)} & $\alpha$ ($^{235}$U $\to$ $^{231}$Th $\to$ $^{227}$Ra)                      & $1.2\times10^{43}$  \\  
$^{238}$U      & \br{99.2742}      & \br{7941.6 $\pm$ 10.4}& \br{$^{230}$Ra (93 m)}   & $\alpha$ ($^{238}$U $\to$ $^{234}$Th $\to$ $^{230}$Ra)                      & $9.3\times10^{50}$  \\  
~              & ~                 & ~                     & ~                        & ~                                                                           & ~                   \\ 
\hline
\multicolumn{6}{l}{$^*$ While $^{231}$Pa is listed in \cite{Mei16} as present in the natural isotopic composition with abundance}  \\
\multicolumn{6}{l}{~~~ $\delta = 100$\%, in fact, it is unstable nuclide with quite short half-life $T_{1/2} = 32760$~y \cite{Fir98}.}  \\
\end{longtable}
\end{footnotesize}

One can see from the Table that all the $2\alpha$ candidates also are potentially unstable in relation to single $\alpha$ decay.
Intermediate $(A-4,Z-2)$ nuclides always live long enough \cite{Fir98} not to imitate $2\alpha$ decay in fast chain of two single
$\alpha$ decays; and in few cases ($^{144,145}$Nd, $^{160}$Dy, $^{167,168}$Er, $^{176}$Yb, $^{207,208}$Pb) $\alpha$ decay of the
intermediate nuclide is energetically forbidden.

Different methods can be used to look for the $2\alpha$ decay, e.g.:

(1) Detection of nuclear recoils which in case of $2\alpha$ decay can have energies higher than those 
in single $\alpha$ decay (depending on the angle between the emitted $\alpha$'s) \cite{Nov79,Ber88};

(2) Si detectors (or some others, e.g. nuclear emulsions) can register two 
$\alpha$ particles emitted from an external source and measure their energies; 
however, samples in this case should be very thin that restricts the mass that can be investigated;

(3) An external source can be placed on e.g. HPGe detector; 
if the daughter nuclide is unstable, one can register characteristic $\gamma$ quanta 
emitted in its decay. In this case, mass of a sample could be a few kilograms; 
however, efficiency of HPGe detectors is on the level of only few percents.
Other possible origins of the daughter nuclides should be also taken into account 
(resulting from e.g. cosmogenic production, or due to fission of U/Th present in 
some amounts in the investigated source);

(4) Very promising is a ``source = detector'' approach, when a mother nuclide 
is embedded in a detector itself as its main component (like W isotopes in CdWO$_4$ 
or $^{209}$Bi in Bi$_4$Ge$_3$O$_{12}$) or as a dopant (like $^{203}$Tl in NaI(Tl)). 
This approach gives possibility to use detectors with big masses 
(up to $\sim ~1$~t) and ensures high efficiency of registration of the 
$2\alpha$ process (practically 100\%). 
Scintillators can be used; however, in this case one cannot expect 
high energy resolution (it will be on the level of few tens or hundreds keV). 
Also, scintillators have different light yields for $\beta$ and $\gamma$ particles 
comparing to those for $\alpha$ particles of the same energy (quenching effect, 
see e.g. \cite{Tre10}). Thus, $2\alpha$ light signal will be quenched and 
shifted to lower energies, where $\alpha$ and $\beta$/$\gamma$ backgrounds are higher. 
Instead, scintillating bolometers, devices able to measure simultaneously light and heat signals 
for the same event (see e.g. \cite{Pir17}), are very promising techniques. 
A signal in the heat channel is not quenched (thus, one will see $2\alpha$ signal 
at $Q_{2\alpha}$ value but not at lower energies). 
In addition, energy resolution in the heat channel is on the level of few keV. 

\section{Theoretical estimation of $T_{1/2}$ for $2\alpha$ decay}

Half-lives for $^{148}$Sm, $^{152}$Gd, $^{156}$Dy, $^{190}$Pt and $^{234}$U were estimated 
in \cite{Ber88}. There is a good agreement between the values of \cite{Ber88} and results 
obtained in this work (within 1 -- 2 orders of magnitude, if to take also into account 
difference in $Q_{2\alpha}$ values known in 1986 and the current $Q_{2\alpha}$'s \cite{Wan17}). 

Calculations of $2\alpha$ decay half-lives for some nuclei in framework of the 
superasymmetric fission model were performed in \cite{Poe85}. The results are 
presented in table \cite{Poe85,Poe89}, and also in graphical form as a logarithm 
of ratio of probability to emit $2\alpha$ to that for single $\alpha$ emission. 
For nuclides presented in Table~1, results are absent. 

To estimate the half-life values for $2\alpha$ emission, we use here semi-empirical 
formulae for cluster decays developed in \cite{Poe02} and applied for emission of 
$^8$Be nucleus. As it is known, $^8$Be is highly unstable decaying to two $\alpha$'s 
with $T_{1/2} \simeq 10^{-16}$~s and $Q = 91.8$~keV \cite{Fir98}. Thus, the energy 
releases in $2\alpha$ decay should be higher on 91.8~keV comparing to decay with 
$^8$Be emission, and corresponding $T_{1/2}$ values should be slightly smaller. 
However, the difference is not big \cite{Poe85} and is acceptable for our aims of $T_{1/2}$ 
estimation. The results for some prospective nuclei are given in Table~1. 
One can see that even for the most promising cases, half-life values are too big for $2\alpha$ 
decay's experimental observation.

\section{First experimental $T_{1/2}$ limit for 2$\alpha$ decay of $^{209}$Bi}

To obtain the first limit on 2$\alpha$ decay of $^{209}$Bi, we can use data 
from the experiment \cite{Mar03} in which single $\alpha$ decay of $^{209}$Bi was 
observed at the first time. In particular, a BGO (Bi$_4$Ge$_3$O$_{12}$) scintillating 
bolometer with mass of 45.7~g was measured during 100~h in this work. 
Because of different light yields for $\beta$ and $\gamma$ particles comparing 
to those for $\alpha$ particles of the same energy, energy spectrum of 
$\alpha$ particles and nuclear recoils is effectively discriminated from much more 
intensive $\beta$ and $\gamma$ background. Because of small range of 
$\alpha$ particles in the BGO crystal, they are absorbed in the crystal 
with practically 100\% efficiency. In the heat channel, we should observe 
a peak for the single $\alpha$ decay of $^{209}$Bi at $Q_\alpha = 3137.3 \pm 0.8$~keV \cite{Wan17}, 
and for the double $\alpha$ decay at $Q_{2\alpha} = 3292.2 \pm 3.5$~keV (see Table~1). 
In the data presented in Fig.~2b of Ref. \cite{Mar03}, one really sees the peak for 
$^{209}$Bi single $\alpha$ decay, while the peak at $Q_{2\alpha}$ is absent. 
In fact, no events are registered in the energy range $3.2 - 3.4$~MeV which fully 
contains the expected $2\alpha$ peak (full width at half maximum, FWHM is $\simeq 15$~keV 
\cite{Mar03}). Thus, we can estimate only $T_{1/2}$ limit for the $^{209}$Bi $2\alpha$ 
decay with the formula:

\begin{center}

$\lim T_{1/2} = \ln 2 \cdot \epsilon \cdot N_{209} \cdot t / \lim S$,

\end{center}

\noindent where $\epsilon$ is the efficiency to detect the expected $2\alpha$ decay 
($\epsilon = 1$), 
$N_{209}$ is the number of $^{209}$Bi nuclei in the 45.7~g Bi$_4$Ge$_3$O$_{12}$ crystal 
($N_{209} = 8.84\times10^{22}$), 
$t$ is the time of measurements ($t = 100$~h), and 
$\lim S$ is the upper limit on the number of events of the effect searched for 
which can be excluded at a given confidence level (C.L.). 
In accordance with the Feldman-Cousins recommendations \cite{Fel98},
for 0 registered events (and with 0 supposed background), $\lim S = 2.44$ at 90\%~C.L. 
Substituting all the values in the formula above, we obtain the following constraint on the 
$^{209}$Bi $2\alpha$ decay:

\begin{center}

$T_{1/2} > 2.9\times10^{20}$~y at 90\%~C.L.

\end{center}

The obtained experimental limit is very far from the theoretical expectations 
presented in Section~3 (for $^{209}$Bi, $T_{1/2} = 4.3\times10^{113}$~y). 

Half-life limit for $2\alpha$ decay of $^{232}$Th, nuclide with quite big energy 
release $Q_{2\alpha} = 8152$~keV and high natural abundance near 100\%, could be derived 
from e.g. measurements \cite{Kim20} of 2~kg ThO$_2$ sample with HPGe detector at 
the Yangyang underground laboratory (Korea). The daughter nuclide $^{224}$Rn 
is unstable and $\beta$ decays further to $^{224}$Fr with $Q = 800$~keV and 
$T_{1/2} = 107$~min \cite{Fir98}. One can look for characteristic $\gamma$ quanta 
emitted in $^{224}$Rn (and $^{224}$Fr) decay.

\section{Conclusions}

Nuclear decay $(A,Z) \to (A-8,Z-4) + 2\alpha$ with simultaneous emission of two alpha particles 
is energetically allowed for near 40\% of known isotopes (1459 from 3436 listed in \cite{Wan17}). 
Among them, 80 nuclides are present in the natural isotopic composition of elements \cite{Mei16}. 
First experimental limit for this kind of radioactivity is given for $^{209}$Bi as 
$T_{1/2} > 2.9\times10^{20}$~y at 90\%~C.L. Theoretical $T_{1/2}$ estimations are calculated for 
the most prospective candidates. However, the calculated $T_{1/2}$ values are very big, 
$10^{33}$~y or more, making prospects for future observation of such processes very pessimistic.

\section*{Acknowledgements}

The work was supported in part by the National Research Foundation of Ukraine 
Grant No. 2020.02/0011. It is my pleasure to thank F.A.~Danevich, V.V.~Kobychev 
and O.G.~Polischuk for useful discussions. 
I am grateful to Yu.N.~Novikov who drew my attention to papers \cite{Nov79,Ber88}.

~

{\it Note added in proofs.}
After appearance of eprint version of this work \cite{Tre21}, very recently $2\alpha$ decay of 
$^{212}$Po and $^{224}$Ra was described in a microscopic framework based on energy density functional 
\cite{Mer21}. The calculated half-lives for two alpha particles emitted in opposite directions 
were found to be 20 (13) orders of magnitude lower for $^{212}$Po ($^{224}$Ra) in comparison 
with those obtained with semi-empirical formulae for $^8$Be emission \cite{Poe02}. 
This gives more hopes for experimental investigation of the process.

\end{document}